\documentclass[12pt, aip,cha,graphicx,amsmath,amssymb,reprint,onecolumn]{revtex4-1}
\usepackage{ amstext, graphicx, amsfonts,amsthm,bm, ulem, color, bbm}
\newcommand{\Geo}{g}

\begin{document}
\title{Active prestress leads to an apparent stiffening of cells through geometrical effects}
\author{E. Fischer-Friedrich$^{1,\ast}$}
\address{$^1$Biotechnology Center, Technische Universit\"at Dresden, Dresden, Germany}

\begin{abstract}
{Tuning of active prestress e.g. through activity of molecular motors constitutes a powerful cellular tool to adjust cellular stiffness through nonlinear material properties. Understanding this tool is an important prerequisite for our comprehension of cellular force response, cell shape dynamics and tissue organisation.
Experimental data obtained from cell-mechanical measurements  often show a simple linear dependence between mechanical prestress and measured differential elastic moduli. While these experimental findings could point to stress-induced structural changes in the material, we propose here a surprisingly simple alternative explanation in a theoretical study. We show how geometrical effects can give rise to increased cellular force response of cells in the presence of active prestress. The associated effective stress-stiffening is disconnected from actual stress-induced changes of the elastic modulus and should therefore be regarded as an apparent stiffening of the material. We argue that  new approaches in experimental design are necessary to  separate this apparent stress-stiffening due to geometrical effects from actual nonlinearities of the elastic modulus in prestressed cellular material. \\ \\
}
%
{*Correspondence: elisabeth.fischer-friedrich@tu-dresden.de}
\end{abstract}

\maketitle
\newpage

\section{Introduction}
Cells need to deform themselves during many physiological processes in the body such as cell division, cell migration or the complex events of morphogenesis. To understand dynamic cell deformation from a physical point of view, we need to quantify cell material properties and its active regulation through the underlying molecular cell biology. 
Cells are a complex viscoelastic material whose  material properties are time-scale dependent and nonlinear beyond strains of few percent \cite{koll11}.
Mechanical stress therefore alters cell-mechanical properties making cells stiffer or softer \cite{koll11, koll09, lang17, stri10, ingb14}.
The material properties of a cell are mainly determined by its cytoskeleton. 
Intriguingly, cells are able to self-tune its mechanical prestress in the cytoskeleton through the action of molecular motors giving the cell a tool at hand to actively regulate its material stiffness through dynamic adjustment of active prestress together with stress-stiffening or stress-softening.
Many studies report stress-stiffening through active stress \cite{koll11, wang02, stam04,fisc16, fern06, jans13}, however opposite findings of cell-softening have also been reported \cite{chan15}.  Stiffness-modulation through the action of motor proteins in actin meshwork's  was further corroborated by {\it in vitro} measurements \cite{mizu07, bend08,koen09}.
It has been suggested that cytoskeletal stress-stiffening may be understood in terms of the nonlinear entropic thermal stretch modulus of  a single polymer \cite{bust94, gard04, stor05} or the stress-induced ``pull-out" of soft bending modes and a transition towards a stretch-dominated regime \cite{ broe11}. As a third option,  load-dependent binding dynamics of actin cross-linkers has been put forward \cite{yao13}. While entropic effects predict a power law scaling of $3/2$ for the resulting differential shear modulus \cite{gard04}, the ``pull-out'' of soft bending modes gives rise to a power law scaling between $1/2$ and $1$ \cite{broe11}.
Furthermore, linear stress-stiffening has been reported for marginal networks \cite{shei12, licu15} {\color{black} and for tensegrity models  \cite{volo00}.}\\
Experimental data obtained from cell-mechanical measurements  show often a linear dependence between  measured differential elastic moduli and active mechanical prestress. This corresponds to a power law with exponent one in dependence of prestress \cite{koll11, wang02, stam04, fisc16}. 
 {\color{black} While these experimental findings could point to complex structural changes of the cytoskeleton through active prestress and accordant modification of its elastic modulus, we propose here a surprisingly simple alternative explanation: pure geometrical effects may give rise to an increased force response and thus an effective stiffening of actively prestressed material. Associated effective stiffening is however independent of actual stress-dependent changes of the elastic modulus as a material parameter of the cytoskeleton in a coarse-grained continuum description. Thus  the associated stiffening will be denoted as `apparent stiffening'.
We will discuss two different geometrical effects i) stress-stiffening through geometrical-coupling and ii) shear-induced nematic alignment. }
In fact, when present, these effects disguise the actual stress-dependence of the elastic modulus and may account partly  for  incoherences in the cell mechanics field such as order of magnitude differences in measured shear moduli and contradicting trends (stiffening or softening) in response to mechanical prestress. 

\section{Apparent stress-stiffening through geometrical coupling}
{\color{black}  
In the following paragraphs, we discuss two simplified examples of cell-mechanical probing to illustrate the general phenomenon of apparent stress-stiffening through geometrical coupling in i)~adherent and ii)~non-adherent cells. 
In our calculations, we will henceforth assume that displacements and strains are small. We will thus follow the approach of linear elasticity theory keeping only terms to first-order  in strain and deformation}. 

\begin{figure*}[h]
\centering
\includegraphics[width=\textwidth]{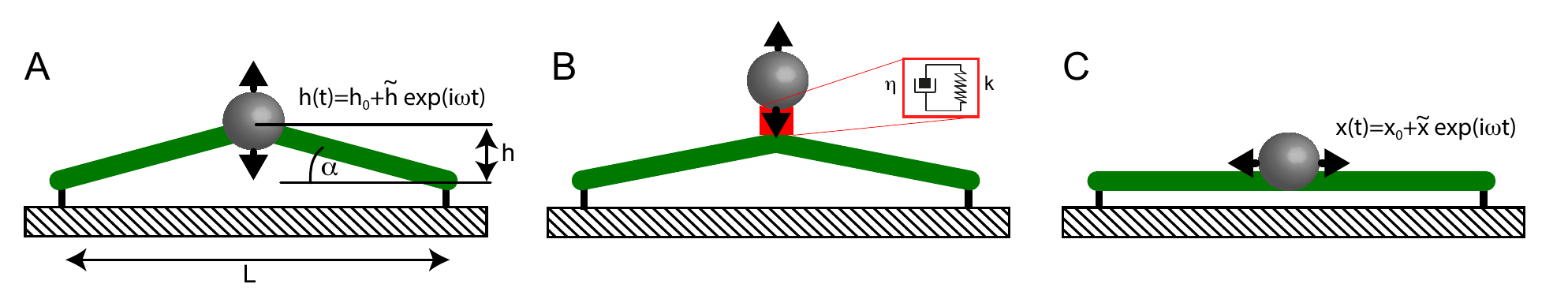}
\caption{ \label{fig:Fig1}
Different setups for rheological probing of a contractile cytoskeletal fibre. The force response of the fiber is affected by the magnitude of inherent active prestress in measurement scenarios A and B. In setup C, contributions of active prestress to force oscillations are avoided.}
\end{figure*}

{\color{black} {\bf Deformation of an adhered model cell.}}  We will first consider the case of a one-dimensional actively prestressed fibre tethered to a substrate as a minimal model of an adherent cell. Consider the experimental scenario sketched in Fig.~\ref{fig:Fig1}A where the fibre center  is oscillated around a mean height with $h(t)=h_0+ \tilde h \exp(i\omega t)$ through a force exerted by an adherent bead. 
The one-dimensional stress in the fibre has a contribution from active prestress $\sigma_{\rm act}$ and a deformation-induced viscoelastic contribution \cite{fisc16}
\begin{equation}
\sigma(t)=\sigma_{\rm act}+ G^\ast(\sigma_{\rm act}, \omega) \epsilon(t) \, ,
\end{equation}
where $G^\ast(\sigma_{\rm act}, \omega)$ is the prestress- and frequency-dependent complex elastic modulus of the fibre, $\epsilon(t)=\tilde\epsilon \exp(i\omega t)$ is the  strain with respect to fibre length changes and $\omega=2\pi f$ is the angular frequency of the applied oscillation.

The measured force signal is 
\begin{equation}
F(t)=\Geo(t)(\sigma_{\rm act}+ G^\ast(\sigma_{\rm act}, \omega)\epsilon(t)) \,  ,
\end{equation}
{\color{black} where $\alpha(t)$ is the angle between the fibre and the substrate (see Fig.~\ref{fig:Fig1}A) and} $\Geo(t)=2 \sin\alpha(t)$ is a time-dependent geometrical factor. 
We make the ansatz
\begin{align}
\Geo(t)&=\Geo_0+\tilde \Geo \exp(i\omega t)\\
F(t)&=F_0+\tilde F \exp(i\omega t) \,  ,
\end{align}
where  $\Geo_0$ and $F_0$ are average values of the geometrical factor and measured force, respectively, while $\tilde \Geo$ and  $\tilde F$ characterise emerging oscillatory changes. Here, $\tilde F$ may take complex values if the respective force oscillations are not in-phase with the imposed oscillations of bead height.  
We find to first order
\begin{align}
F(t)=F_0+\big(\tilde \Geo \sigma_{\rm act}+\Geo_0 G^\ast(\sigma_{\rm act}, \omega) \tilde\epsilon\big)\exp(i\omega t)  \, .
\end{align}
Force oscillation amplitudes thus contain a contribution $\tilde \Geo \sigma_{\rm act}$  proportional to the active prestress  if the oscillation amplitude of the geometrical factor $\Geo$ is does not vanish. We will denote this effect as geometrical coupling of active stress into the force signal. 
In a conventional rheological analysis, that disregards the presence of active prestress, the force amplitude is assumed to scale with the complex elastic modulus of the system for a given strain.
Therefore,  the effective elastic modulus would be determined to scale as
\begin{equation}
G_{\rm eff}^{\ast}(\sigma_{\rm act}, \omega)\propto \frac{1}{ \Geo_0}\frac{\tilde \Geo}{ \tilde\epsilon} \sigma_{\rm act}+G^\ast (\sigma_{\rm act}, \omega)  \,  .  \label{eq:Gmeas}
\end{equation}
A short calculation yields {\color{black} (see Supporting Material)}
\begin{equation}
\frac{1}{ \Geo_0}\frac{\tilde \Geo}{ \tilde\epsilon} =\frac{L^2}{4h_0^2}  \,  ,
\end{equation}
 where $L$ is the fibre length in a straight conformation (see Fig.~\ref{fig:Fig1}A).
Thus, the effective shear modulus shows apparent linear stress-stiffening with a slope proportional to $L^2/4h_0^2$ and an additional active contribution  $ \frac{1}{ \Geo_0}\frac{\tilde \Geo}{ \tilde\epsilon} \sigma_{\rm act}$.
In particular, for sufficiently small values of mean height $h_0$, the active contribution starts to override the contribution of the actual complex elastic modulus in the measurement and the slope of the apparent stress-stiffening can in principle become arbitrarily large.  \\
So far, the examples of geometrical coupling of  active prestress have contributed only a constant contribution to the effective storage modulus of the system because the deformation and the geometrical factor were in phase. However, in general, out-of-phase oscillations of the geometrical factor are possible e.g. in the case of a prestressed material coupled to a viscoelastic connector (see Fig.~\ref{fig:Fig1}B and Supporting Material). 
If an attached bead is deflected horizontally along the fiber (see Fig.~\ref{fig:Fig1}C), active contributions to the force oscillations are avoided and we have $G^{\ast}_{\rm eff}(\sigma_{\rm act}, \omega)=G^{\ast}(\sigma_{\rm act}, \omega)$. 

\begin{figure*}[h]
\centering
\includegraphics[width=\textwidth]{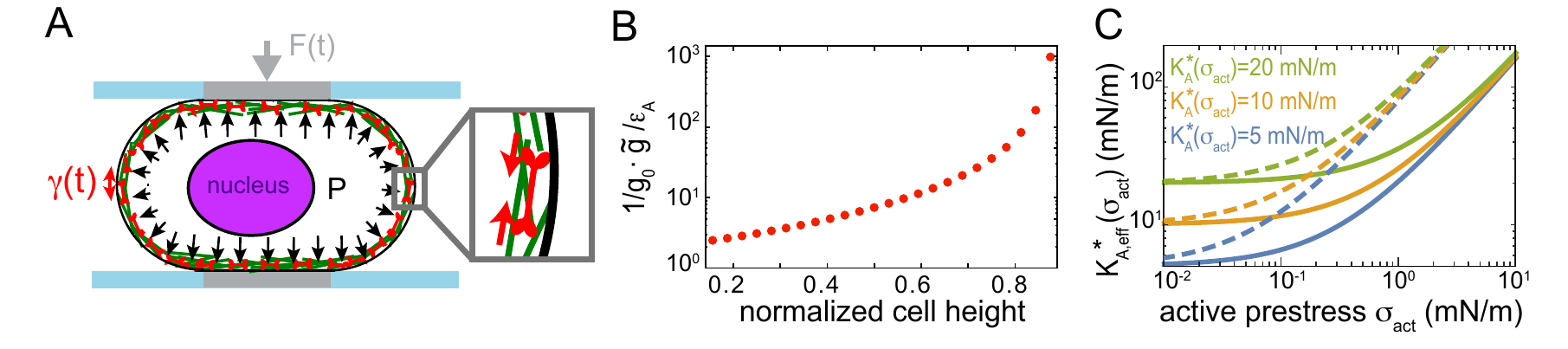}
\caption{ 
Oscillatory parallel plate confinement of non-adherent cells. A) Cells are uniaxially compressed between parallel plates oscillating around a mean confinement height $h_0$. 
Cortical contractility of the cell is balanced by an internal hydrostatic pressure. During the oscillation cycle, cell contact area and curvature of the free cell surface changes as well as the internal pressure and the mechanical stress in the outer cortex-membrane layer. 
B) Apparent stress-stiffening slope  $1/\Geo_0\cdot \tilde \Geo/\tilde\epsilon_A$ in dependence of average cell confinement height, normalized by the cell diameter in a spherical conformation.
C) Apparent modulus $K_{A,\rm eff}^\ast (\sigma_{\rm act})$  calculated according to the r.h.s. of Eq.~\ref{eq:KAmeas}  for a cell confined to a normalized cell height of  $0.65$ (solid lines) and $0.8$ (dashed lines). The actual area dilation modulus $K_A^\ast (\sigma_{\rm act})$ was chosen to be constant and thus stress-independent (blue: $5$~mN/m, yellow: $10$~mN/m, green: $20$~mN/m). Parameters were motivated by measurements of mitotic cells \cite{fisc16}.
\label{fig:Fig2}
}
\end{figure*}

{\bf \color{black} Deformation of a non-adherent model cell.} As a second example, we consider a non-adherent cell that is being probed in a parallel plate assay subject to oscillatory height changes $h(t)=h_0+ \tilde h \exp(i\omega t)$ (see Fig.~\ref{fig:Fig2}A). Accordant measurements were for instance performed on mitotic cells in Ref.~\cite{fisc16}. We will regard the cell as a liquid-filled pressurized shell where the shell material is constituted by the plasma membrane and the neighboring actomyosin cortex.  To keep our calculations simple, we will assume that the deformation-induced stress in the cortex-membrane layer is a spatially homogeneous, isotropic, in-plane stress governed by an area dilation modulus and an active mechanical prestress $\sigma_{\rm act}$. 
Furthermore, we will assume that viscous contributions of the cytoplasm are negligible \cite{fisc16}. With these simplifications,  Laplace's law $\Delta p= 2 H \sigma$ holds at all times for the free cell surface area, where $H$ denotes the mean curvature of the cell surface, $\sigma$ the two-dimensional mechanical  stress  in the cortical shell and $\Delta p$ a balancing pressure excess  in the cytoplasm \cite{fisc14}.
Analogous to the previous example, we have 
\begin{equation}
F(t)=\Geo(t)\left(\sigma_{\rm act}+ K_A^\ast(\sigma_{\rm act}, \omega) \epsilon_A(t)\right)   ,
\end{equation}
where $\Geo(t)$ is again a geometrical factor,  $K_A^\ast(\sigma_{\rm act}, \omega)$ is the prestress- and frequency-dependent complex area dilation modulus and $\epsilon_A(t)=\tilde\epsilon_{A} \exp(i\omega t)$ is the corresponding surface area shear with amplitude $\tilde\epsilon_{A}$.
If the cell is not adhering to the contacting plate, we have a vanishing contact angle and thus according to Laplace's law  $\Geo(t)=2 A_{\rm con}(t)  H(t)$ where $A_{\rm con}(t)$ denotes the time-variable contact  area of the cell-plate interface.
Due to Laplace's law, the oscillatory variations of the geometrical factor are in phase with surface area oscillations such that $\tilde\Geo$ is real. 
We  find analogous to Eq.~\ref{eq:Gmeas}
\begin{align}
\tilde F\propto K^{\ast}_{A, \rm eff}(\sigma_{\rm act}, \omega)\propto \frac{1}{ \Geo_0}\frac{\tilde \Geo}{ \tilde\epsilon_{A}} \sigma_{\rm act} +K_A^\ast(\sigma_{\rm act}, \omega) \, .
\label{eq:KAmeas}
\end{align}
Thus, the effective modulus $K^{\ast}_{A, \rm eff}(\sigma_{\rm act}, \omega)$ contains a term linear in active stress giving rise to an (apparent) linear stress-stiffening in addition to possible actual stress-stiffening of the material given by the explicit dependence of  $K_A^\ast(\sigma_{\rm act}, \omega)$ on $\sigma_{\rm act}$. To illustrate the magnitude of this apparent linear stress-stiffening, we estimated the stiffening-slope $1/\Geo_0\cdot \tilde \Geo/\tilde\epsilon_A$  for a cell  subject to different degrees of confinement as described in Ref.~\cite{fisc16}. It is noteworthy that the  stiffening slope  $1/\Geo_0\cdot \tilde \Geo/\tilde\epsilon_A$ is independent of the strain amplitude $\tilde\epsilon$ and may take arbitrarily large values as it diverges for {\color{black} large cell confinement heights} (see Fig.~\ref{fig:Fig2}B). 

{\color{black} In conclusion, the two examples presented above show that geometrical coupling may give rise to an apparent linear stress-stiffening during cell-mechanical measurements with stiffening slopes that depend crucially on the chosen measurement setup. Therefore, the associated stiffening slopes are no material parameters.}

\section{Apparent stress-stiffening through shear-induced  nematic alignment}
\begin{figure*}[b]
\centering
\includegraphics[width=\textwidth]{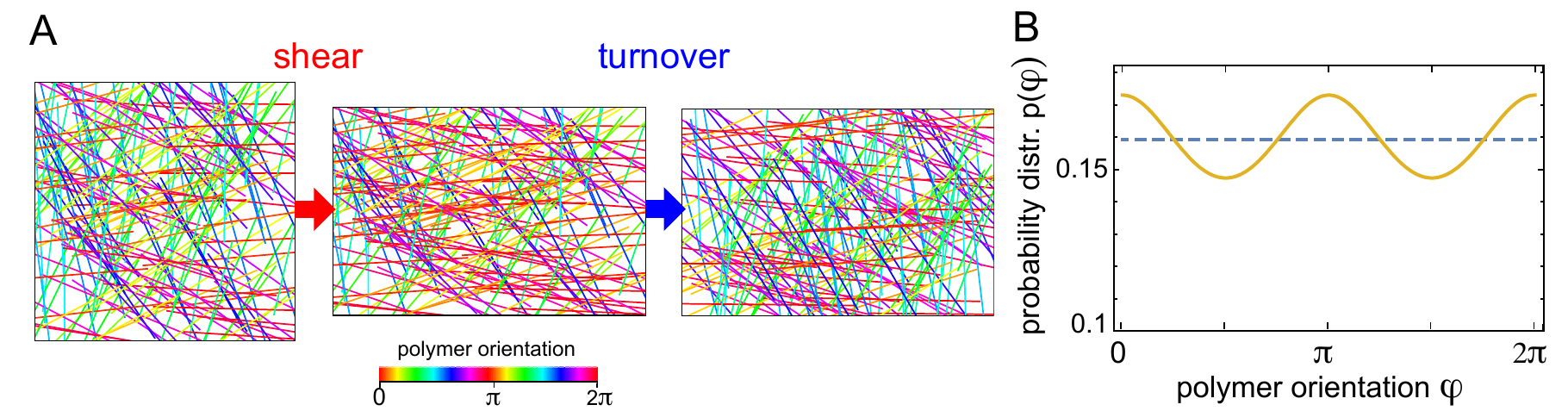}
\caption{ \label{fig:Fig3}
Nematic alignment through shear. A) Randomly oriented polymer rods (left panel) are reoriented by a shear deformation with $\lambda=0.2$. Shear deformations induce nematic alignment of biopolymers leading to a deformation-resistance proportional to active prestress  (middle panel). Turnover reestablishes the original orientation distribution of polymers  (right panel). The color legend indicates the orientation angle $\varphi$ of polymers of different color.
B) Polymer distribution before (blue) and after (yellow)  shear deformation with $\lambda=0.2$. }
\end{figure*}

{\color{black} So far, we have considered examples of cell-mechanical probing where no shear deformations have been involved. Typically, shear deformations do however play a role and give rise to an additional form of apparent linear stress-stiffening through geometrical effects. Stiffening through fibre-alignment and active prestress has  already been discussed in previous papers \cite{stam02, ingb14}. We reformulate it here in the continuum mechanics framework of active gel theory. Furthermore, we incorporate as a new aspect the effect of filament turnover.
Nematic alignment of cytoskeletal polymers  has been discussed before in the case of viscous flows in active gels \cite{reym16} but has, to our knowledge, not yet been explicitly connected  to stress-stiffening in cells.
  
In the following, we  discuss how shear deformations change the nematic order of a polymer network leading to an apparent linear stiffening in the presence of active prestress. }
 For simplicity, we discuss  a two-dimensional material. In this case, the nematic order tensor is  defined as
\begin{align}
{\bf Q}=\int_0^{2\pi} 
\begin{pmatrix}
\cos(\varphi)^2-\frac{1}{2} & \cos(\varphi)\sin(\varphi) \\
\cos(\varphi)\sin(\varphi) & \sin(\varphi)^2 -\frac{1}{2}
\end{pmatrix} p(\varphi) {\rm d}\varphi  \label{eq:NematicOrder} \, ,
\end{align}
where $ p(\varphi)$ is the probability distribution of finding a polymer with an orientation of polar angle $\varphi$. We will assume that  the network is isotropic in the reference configuration and $ p(\varphi)=1/(2\pi)$. Thus, before deformation, ${\bf Q}=0$. After application of an affine shear deformation to an area element, this orientation distribution will be changed (see Fig.~\ref{fig:Fig3}A). To illustrate that, consider a small homogeneous shear whose principle axes are, without loss of generality, in x-y-direction. The associated  strain tensor reads
\begin{align}
{\boldsymbol\epsilon}=
\begin{pmatrix}
\lambda & 0\\
0 & -\lambda
\end{pmatrix}  \, .
\end{align}
 In this case, a polymer orientation $\varphi$ will on average be realigned by the mapping \hbox{$\varphi\rightarrow \varphi - 2 \lambda \cos(\varphi)\sin(\varphi)$} to first order in $\lambda$, leading to a new probability distribution \hbox{$p^\prime(\varphi)=1/2\pi+\cos(2\varphi)\lambda/\pi$}. In this way,  polymer orientations along the x-axis become more likely  (see Fig.~\ref{fig:Fig3}). As a consequence, the active prestress becomes anisotropic. Using Eq.~\ref{eq:NematicOrder}, the nematic order tensor is now to first oder in strain ${\bf Q}=\boldsymbol\epsilon/2$. Following active gel theory \cite{pros15}, we make the ansatz 
 \begin{align}
 \sigma^{\rm act}({\boldsymbol\epsilon})=\zeta_1\Delta\mu\, {\bf 1}+\zeta_2 \Delta\mu {\bf Q}({\boldsymbol\epsilon})
 \end{align}
 for active tension in the cytoskeleton, where $\Delta\mu$ denotes the difference between the chemical potentials of ATP and its hydrolysis products.
 If the material is elastic with prestress-dependent shear modulus $G(\sigma_{\rm act})$, the overall traceless stress  after a shear deformation is thus
 \begin{align}
2\left(\frac{\zeta_2}{8\zeta_1}\sigma^{\rm act}_{ii}(0)+G(\sigma_{\rm act})\right)\boldsymbol\epsilon  \,  .
 \end{align}
Therefore, in an analysis of the force response that does not compensate for the influence of active prestress, the effective shear modulus would thus be identified as 
 \begin{equation}
 G_{\rm eff}(\sigma_{\rm act})=\frac{\zeta_2}{8\zeta_1}\sigma^{\rm act}_{ii}(0)+G(\sigma_{\rm act}) .
\end{equation}
Since the cytoskeleton turns over through depolymerization and repolymerization processes, induced nematic alignment may be lost over time with the original polymer orientation distribution being reestablished. 
In the following, we will adapt our above considerations on shear-induced nematic alignment to a viscoelastic material. Adopting the scheme in Reymann {\it et al.}\cite{reym16}, we will assume that the material turnover takes place on a time scale $\tau_{to}$ and that a Maxwell-type dynamics governs the time dependence of the the nematic order tensor such that $\dot {\bf Q}=\dot {\boldsymbol \epsilon}/2- {\bf Q}/\tau_{to}$. In Fourier space, we obtain correspondingly 
\begin{align}
{\bf Q}(\omega)&=\frac{i\omega}{(i\omega+1/\tau_{to})}\frac{\boldsymbol\epsilon(\omega)}{2} \,  .
\end{align}
Thus, 
 \begin{equation}
 G_{\rm eff}^\ast(\sigma_{\rm act}, \omega)=\frac{\zeta_2}{8\zeta_1}\sigma^{\rm act}_{ii}(0) \frac{i\omega}{(i\omega+1/\tau_{to})}+G^\ast(\sigma_{\rm act},\omega) \, ,
  \end{equation}
which includes an apparent linear stress-stiffening term in addition to the actual shear modulus.
For a homogeneous shear in a three-dimensional material,  a corresponding calculation gives 
 \begin{equation}
 G_{\rm eff}^\ast(\sigma_{\rm act}, \omega)=\frac{\zeta_2}{{\color{black} 15}\zeta_1}\sigma^{\rm act}_{ii}(0) \frac{i\omega}{(i\omega+1/\tau_{to})}+G^\ast(\sigma_{\rm act},\omega) \, .
\end{equation}
 The expected  stress-stiffening slope of the storage modulus due to nematic alignment  is proportional to $\zeta_2/\zeta_1$. As the parameters $\zeta_1$ and $\zeta_2$ are {\it a priori} not known, the magnitude of the associated stress-stiffening slope is not entirely clear. A simple consideration gives an order of magnitude estimate; if polymers of different orientation in the cytoskeletal network do not interact to generate active prestress, active prestress is proportional to the averaged tensor product of filament orientations and we have  $\zeta_1=\zeta_2/2$ in 2D ($\zeta_1=\zeta_2/3$ in 3D). Thus, the associated stress-stiffening slope would be $1/2$ (in 3D: ${\color{black}3/5}$). 
It is noteworthy that shear-induced nematic alignment will influence material properties also in the absence of active prestress. However, in this case, effects of nematic alignment and emergent anisotropies on the force response are second order in strain {\color{black} and thus negligible for small deformations \cite{land86}}. Only in the presence of active prestress, these effects become first order {\color{black} and non-negligible}.

\section{Discussion}
We have presented theoretical considerations that predict cellular force response to scale linearly with active mechanical prestress even in the absence of actual nonlinear material properties (i.e. if $dG^\ast(\sigma_{\rm act})/d\sigma_{\rm act}=0$).  The resulting apparent stress-stiffening roots in an increase of measured force amplitudes due to  geometrical coupling of active prestress and/or shear-induced nematic alignment of the underlying polymer network during dynamic deformation of the cellular material. 
{\color{black} In particular, geometrical coupling may change effective cell moduli in the presence of cytoskeletal contractility (= active prestress) by orders of magnitude  depending on the choice of measurement parameters (see Fig.~\ref{fig:Fig2}). This phenomenon may contribute to the puzzling diversity of cell mechanical moduli reported in the literature \cite{hoff06}. 
Stress-stiffening through active prestress in cells has been shown before in several experimental studies with  mainly linear stiffening trends and slopes between 0.2-45 (see Table~\ref{tab:Tabl1}) \cite{stam02, wang02, stam04, koll11b, schl15, fisc16}. An influence of geometrical coupling or shear-induced nematic alignment on previously reported stress-stiffening slopes  is conceivable. \\
}
\begin{table*}
{\color{black}
\begin{tabular}{| p{3.5cm}| p{5cm}| p{3cm} |  p{4cm}| }
\hline
{\bf Reference} & {\bf Method} & {\bf Approximate stiffening slope} & {\bf Cells}\\
\hline
Wang {\it et al.}\cite{wang02}, Stamenovi\'c {\it et al.}\cite{stam02} & magnetic twisting cytometry and traction force microscopy & $\approx 0.2$ & adhered HASM cells \\
\hline
Stamenovi\'c {\it et al.}\cite{stam04} &magnetic twisting cytometry and traction force microscopy&   \hbox{$\approx 3$~to~$6.6$, for}  $f\!=0.1$ to $1000$~Hz& adhered HASM cells \\
\hline
Kollmansberger {\it et~al.}\cite{koll11b} & magnetic tweezer &  $1.68$ & adhered, several cell lines\\
\hline
Fischer-Friedrich {\it et~al.}\cite{fisc16} & AFM cell confinement &  \hbox{33 and 45 for}  $f\!=0.1$ and $1$~Hz & suspended, mitotic HeLa cells\\
\hline
\end{tabular}
}
\caption{\label{tab:Tabl1} Summary of previously measured stress-stiffening of cells in response to active prestress.}
\end{table*}
If the geometrical factor of geometrical coupling is changing in phase with applied strain during the measurement, geometrical coupling gives merely rise to {\color{black} a frequency-independent addition} to the effective storage modulus. If the actual storage modulus approaches zero at small frequencies, the active prestress contribution may thus be identified as  $G_{\rm eff}^\prime(\sigma_{\rm act}, \omega\rightarrow 0)$. If the coupling geometrical factor can be estimated, this offers the chance to measure both the complex elastic modulus and the active prestress jointly from one rheological measurement: {\color{black} To this end, the oscillatory force signal needs to be divided by the time-dependent geometrical factor to obtain a stress signal for rheological analysis (see also Fischer-Friedrich {\it et al}. \cite{fisc16}). The resulting modulus is then the actual elastic modulus of the material. On the other hand, dividing the baseline force $F_0$ by the time-averaged geometrical factor $g_0$ yields the active prestress. }
In some ideal measurement setups, geometrical coupling can be entirely avoided (see Fig.~\ref{fig:Fig2}C). However, such setups are not easily experimentally realised. 

{\color{black} Shear-induced nematic alignment  generates a frequency-dependent apparent stiffening of actively prestressed material. However, resulting apparent stress-stiffening is independent of the measurement method  and thus a true material parameter. Thus, disentangling the active stiffness contribution from the actual passive modulus might not even considered to be necessary.
Many measurements report cell rheology  which exhibits diverging moduli at high frequencies and short time scales \cite{koll11, trep08}. 
As active stiffness contributions through shear-induced nematic alignment are bounded by $\zeta_2/\zeta_1\sigma^{\rm act}_{ii}(0)$, they are expected to give negligible contributions to an effective elastic modulus at large frequencies.}

It is likely that the geometrical stress-stiffening effects described here may also contribute to nonlinear mechanics of cells in the presence of {\it external} mechanical prestress. However, exertion of a constant external prestress to viscoelastic cellular material leads in general to i) large deformations (e.g. through a constant strain rate in a rheometer) and ii) large inherent anisotropies of the material. Both effects further  complicate a theoretical description considerably.
{\color{black} Our results indicate that effective elastic moduli may become entirely determined by active prestress  for sufficiently large values of prestress. In the context of external prestress, this could provide an alternative explanation for the recent finding of externally prestressed collagen networks whose shear moduli were found to be collagen concentration independent at sufficiently large mechanical loadings \cite{licu15}.}

{\color{black} In summary, our findings show that geometrical effects may disguise the actual stress-dependence of cellular  elastic moduli during cell-mechanical measurements in the presence of active prestress. New approaches in experimental design are necessary  to separate apparent stiffening from actual  stress-induced stiffening or softening of cells. }

\section*{Acknowledgements}
I thank Benjamin Friedrich, Stefan M\"unster and Ben Fabry for discussions on the topic and critical reading of the manuscript. 


\section*{Supporting Material}
\subsection*{Geometrical coupling with phase-shifted, frequency-dependent active force contributions}
In the first and second example in the main text, geometrical coupling of  active prestress has contributed only a frequency-independent addition to the measured storage modulus of the system. In the following, we will present an example where geometrical coupling gives rise to a complex-valued, frequency-dependent addition to the effective elastic modulus of the system for the case of a deflected prestressed fibre with a viscoelastic connector between the bead and the fiber (see Fig.~1B, main text). The bead is deflected in a vertical manner.
The combined system has a (complex) spring constant 
$$
k_{comb}=1/(1/k_{conn} +1/k_{fibre}),
$$
where $k_{conn}$ and $k_{fibre}$
are the effective (complex) spring constants of the connector and the fibre with respect to vertical deflection.
One finds
$$
k_{fibre}(\sigma_{\rm act}, \omega)=16 h_0^4/(4 h_0^2 + L^2)^{3/2} G^{\ast}_{\rm meas}(\sigma_{\rm act}, \omega),
$$
where $G^{\ast}_{\rm meas}$ equals the r.h.s of formula~(6), main text.
Thus,  in the stiffness of the combined system $k_{comb}$, the active term in $G^{\ast}_{\rm meas}$   contributes in general also to the imaginary part of the system and is frequency-dependent.

{\color{black}
\subsection*{Determining the geometrical factor for the case of a deflected prestressed cytoskeletal fibre}
In the main text, we where discussing the example of geometrical coupling of active stress in a prestressed cytoskeletal fibre tethered at its end points at a distance $L$  (see Fig.~1A, main text). 
Height oscillations are imposed on the fibre center with $h(t) = h_0+ \tilde h \exp(i\omega t)$. Here, we derive the factor of geometrical coupling $1/ \Geo_0 \cdot \tilde \Geo/{\tilde \epsilon}$ for this specific example.

We make a perturbation calculation determining the fibre oscillation dynamics up to first order in height amplitude $\tilde h$. We make the following expansions of the dynamic fibre length $l(t)$ and the dynamic angle $\alpha(t)$ between the substrate and the fibre
\begin{eqnarray}
l(t)& =& l_0+\tilde l \exp(i\omega t)+\mathcal{O}(\tilde h^2), \notag\\
\alpha(t)& =&\alpha_0+\tilde \alpha \exp(i\omega t)+\mathcal{O}(\tilde h^2).\notag
\end{eqnarray}
Using the geometrical relations $\sin(\alpha(t)) =h(t)/(l(t)/2)$ and  $h(t) /(L/2)   =\tan(\alpha(t))$, we obtain
\begin{eqnarray}
l_0 & =&\sqrt{4 h_0^2 + L^2}, \notag\\
 \alpha_0  &= &\arctan(2 h_0/L),  \notag\\
 \tilde\alpha & =& (2 \tilde{h} L)/(4 h_0^2 + L^2),  \notag\\
   \tilde{l} & =& (4 h_0 \tilde{h})/\sqrt{4 h_0^2 + L^2}. \label{eq:tildeRelations}
 \end{eqnarray}
We can thus calculate the time variation of the geometrical factor $g(t)=2 \sin(\alpha(t))$ to first order
\begin{eqnarray*}
g(t) &\equiv& g_0+\tilde g\exp(i\omega t) +\mathcal{O}(\tilde h^2)\\
& =& 2\sin(\alpha_0)+2 \cos(\alpha_0)\tilde\alpha\exp(i\omega t) +\mathcal{O}(\tilde h^2)\\
&=& \frac{2\tan(\alpha_0)}{\sqrt{1+\tan(\alpha_0^2)}}+ \frac{2\tilde\alpha}{\sqrt{1+\tan(\alpha_0^2)}}\exp(i\omega t) +\mathcal{O}(\tilde h^2) \\
& =& \frac{4h_0}{\sqrt{4 h_0^2+L^2}}+ \frac{4\tilde h L^2 }{(4 h_0^2 + L^2)^{3/2}}\exp(i\omega t) +\mathcal{O}(\tilde h^2),
 \end{eqnarray*}
where we have used  Eqn.~\eqref{eq:tildeRelations} in the last transformation step.
We conclude that $g_0=\frac{4h_0}{\sqrt{4 h_0^2+L^2}}$ and $\tilde g=\frac{4\tilde h L^2 }{(4 h_0^2 + L^2)^{3/2}}$.
The strain amplitude of the fibre is $\tilde\epsilon=\tilde l/l_0 $.  We thus  obtain  the coupling geometrical factor as
\begin{eqnarray*}
\frac{1}{g_0}\frac{\tilde g}{\tilde \epsilon}&=& \frac{1}{\frac{4h_0}{\sqrt{4 h_0^2+L^2}}} \frac{\frac{4\tilde h L^2 }{(4 h_0^2 + L^2)^{3/2}}}{\frac{\tilde l}{ l_0}}\\
&=& \frac{L^2}{4h_0^2}.
 \end{eqnarray*}

\end{document}